\documentclass{emulateapj}
\usepackage{graphicx, amsmath, amssymb, amsthm}

\makeatother

\newcommand\etal{{\it et~al.~}}

\begin{document}
\title{Three-Dimensional Dynamical Instabilities in Galactic Ionization Fronts}
\author{Daniel J. Whalen\altaffilmark{1} \& Michael L. Norman\altaffilmark{2}}
\altaffiltext{1}{Applied Theoretical Physics (X-2), Los Alamos National
Laboratory}
\altaffiltext{2}{Center for Astrophysics and Space Sciences,
University of California at San Diego, La Jolla, CA 92093, U.S.A.
Email: dwhalen@cosmos.ucsd.edu}

\begin{abstract} 

Ionization front instabilities have long been of interest for their suspected role in 
a variety of phenomena in the galaxy, from the formation of bright rims and 'elephant
trunks' in nebulae to triggered star formation in molecular clouds.  Numerical treatments
of these instabilities have historically been limited in both dimensionality and input physics,
leaving important questions about their true evolution unanswered.  We present the first
three-dimensional radiation hydrodynamical calculations of both R-type and D-type ionization
front instabilities in galactic environments (i.e., solar metallicity gas).  Consistent with linear stability analyses of
planar D-type fronts, our models exhibit many short-wavelength perturbations growing at early times 
that later evolve into fewer large-wavelength structures.  The simulations demonstrate that 
both self-consistent radiative transfer and three-dimensional flow introduce significant morphological 
differences to unstable modes when compared to earlier two-dimensional approximate models.  
We find that the amplitude of the instabilities in the nonlinear regime is primarily determined by the efficiency 
of cooling within the shocked neutral shell.  Strong radiative cooling leads to long, 
extended structures with pronounced clumping while weaker cooling leads to saturated modes 
that devolve into turbulent flows.  These results suggest that expanding H II regions may 
either promote or provide turbulent support against the formation of later generations of
stars, with potential consequences for star formation rates in the galaxy today.

\end{abstract}


\section{Introduction}

This is the first of two papers that study ionization front (I-front) instabilities in the 
nonlinear regime using three-dimensional radiation hydrodynamical simulations.  In this paper
we consider I-fronts propagating in a gas of solar metallicity, appropriate to galactic
massive star forming environments.  In a second paper, we simulate I-front propagation in
primordial gas with low metallicities relevant to star formation at high redshifts.  We
demonstrate that the cooling properties of the gas have a large influence on how I-front
instabilities manifest.

These instabilities have been the subject of numerous analytical and numerical studies for the
past fifty years, primarily for their suspected role in the formation of inhomogeneities observed in galactic 
emission nebulae since the 1930's \citep{kh58,ax64,gsf96,fet98,rjw99,rjw02}.  Initial corrugations in the
fronts due to ambient density fluctuations are thought to elongate into ionized fingers extending ahead of
the front that can evolve into the 'elephant trunks', bright rims, and cometary globules commonly seen in the 
galaxy today.  Dynamical instabilities in H II regions are of direct relevance to the environments of massive
stars \citep{fhy03} and gamma-ray burst progenitors in general, possibly acting in concert with stellar winds 
to drive clumping of gas that later become susceptible to gravitational instabilities and collapse.  
Perturbations in the ionization front shocks of OB associations may also come into play in scenarios of 
triggered star formation within molecular clouds \citep{det07}.  Historically, studies of I-front instabilities
have progressed in both dimensionality and physics along two lines: unstable modes in R-type fronts and in 
D-type I-fronts (see \citet{o89} for a review of modern I-front nomenclature and \citet{wn06} for numerical 
examples in one dimension).  

\subsection{Shadowing Instabilities in R-Type Fronts}

R-type I-fronts propagate supersonically through media with minimal hydrodynamical coupling to the gas.
The stability of this type of front to encounters with modest density perturbations was examined by \citet{v62}
and \citet{na67}.  \citet{na67} found the fronts to be weakly unstable but were unable to follow their evolution into 
D-type structures.  \citet{brt89} studied the flash ionization of diffuse clumps in which the front does not
revert to D-type (the 'cloud-zapping regime').  He concluded that the clumps evaporate without forming new 
inhomogeneities in their wake but could not determine the downstream fate of the front itself.  Several 
groups have investigated the shadowing of I-fronts by dense clumps \citep{met98,cet98,sok98}, with particular 
focus on the evolution of the neutral tails behind the clumps.  Most recent in this vein are the studies by 
\citet{shp04} of cosmological minihalo photoevaporation by external Population III and quasar ionizing sources at high
and intermediate redshifts.  The aim of these simulations was to assess the role of minihalos as sinks of 
ionizing UV photons during cosmological reionization.  Shadows rather than instabilities arose in these 
models because the obstructions trapped the I-fronts as they evaporated.
 
\citet{rjw99} revisited instability modes in R-type fronts, in this instance following their transition  
to D-type far downstream of the original perturbation.  Their analysis revealed that the overdensity indents 
a planar R-type front with a dimple that first elongates but then saturates (Fig 1 in \citet{rjw99}).  
The ionizing photons are oblique to the sides of the dimple but incident to its tip.  Since the flux is 
lower at the sides than at the tip, as the front approaches the Str\"{o}mgren position the shock breaks 
through the front along the wall of the dimple before its tip.  Still R-type, the tip races forward, 
elongating the dimple into a nonlinear jet instability when the transition to D-critical is complete.  
If the shocked gas efficiently cools, the thin-shell hydrodynamic instability appears and fragments the gas into 
clumps with densities hundreds of times greater than the ambient medium.

\subsection{Dynamical Instabilities in D-Type Fronts}

D-type I-fronts move supersonically with respect to the upstream neutral gas, but subsonically (or sonically)
with respect to the downstream ionized gas.  Consequently, D-type fronts are preceded by shocks in the neutral
gas.
Soon after the modern nomenclature for I-fronts was established, \citet{kh58} and \citet{ax64} examined the 
stability of perturbations in planar weak D-type fronts.  They discovered that recombinations to the ground 
state within ionized fingers advancing ahead of the front effectively dampened their further growth upon 
reaching lengths of a few ${n_{i}}^{-1}$ pc, where n$_{i}$ is the electron density in the postfront gas.  
This is the product of the sound speed of the ionized gas and its recombination time; in galactic environments 
such corrugations are initially short wavelength and saturate quickly.  As the fingers lengthen, photons must 
cross more neutrals resulting from recombinations, which attenuate the ionizing flux advancing the tip of the 
disturbance.

Instabilities that do exhibit strong growth were later discovered in D-type ionization fronts, but only in
conjunction with pre-existing flow instabilities \citep{v83,cap73,bnd81}.  In general, expanding spherical 
shocks are prone to a variety of dynamical instabilities that can result in their fragmentation and breakup.  
Radiating shocks are subject to the Vishniac (or thin-shell) instability in either uniform densities or 
gradients, while Rayleigh-Taylor instabilities can arise in adiabatic or radiating flows if they accelerate 
(which occurs in radial power-law density gradients steeper than r$^{-2}$).  

The Vishniac mechanism can be visualized as follows (Fig 1 of \citet{mn93}): in the frame of the shock, the 
inflow ram pressure is unidirectional while the ionized postshock pressure is isotropic.  If a density 
fluctuation is advected across the shock, small transverse velocities arise across the face of the shock 
that create overdensities and underdensities along adjacent lines of sight from the center of the shell. 
The postshock gas then protrudes through the underdense regions, dimpling the shock into peaks and valleys 
and forcing even greater transverse flows that further bunch gas across adjoining lines of sight.  However, 
the higher pressures in the troughs eventually reverse the transverse flows, resulting in oscillatory ripples 
with modest growth rates across the face of the shock commonly known as overstabilities.   

Both thin-shell overstabilities and Rayleigh-Taylor instabilities are exacerbated when driven by an 
ionization front \citep{sp54,f54,gu79,gsf96,rjw00,miz05,miz06}.  When perturbations appear in the shell, the 
radiation preferentially 
advances into the lower optical depths of the underdensities, reinforcing the transverse velocity gradients
that originally created them in a repeat of the cycle described above (Fig 1 of Garcia-Segura \& Franco).  
What distinguishes unstable modes in blast waves or wind-blown bubbles from those in D-type fronts is their 
amplitude:  in extreme cases the radiation may escape through cracks in the shock, driving even greater 
hydrodynamical growth \citep{gsf96,fet98,fhy03}.  The violence of flow instabilities in general is largely 
governed by the cooling rates of the shocked gas.  Efficient cooling collapses the shocked neutral gas into 
a thin, dense cold shell more prone to breakup than a thick semi-adiabatic shock.  We note that a class of 
long-wavelength unstable surface modes in weak D-type fronts has recently been uncovered by analytical and 
numerical work done by \citet{sy97} and \citet{rjw02} that are independent of any other flow instabilities, 
but we do not consider them here.

A key question is how the dimensionality of ionized flows dictates the nature of perturbation growth in 
R-type and D-type fronts, given that published simulations have been confined to two dimensions.  Furthermore, 
instability growth is mediated by I-front transitions in prefront shocked gas that mandate radiation 
hydrodynamics rather than the approximations heretofore performed \citep{gsf96,det07,yet06}.  
In order to explore these issues we have performed the first three-dimensional radiation hydrodynamical 
calculations of dynamical instabilities in D-type fronts \citep{fet98} and shadowing instabilities in R-type 
fronts \citep{rjw99} with metal-line cooling.  These models have been made possible by recent upgrades to the 
ZEUS-MP reactive flow physics code \citep{wn06} (hereafter WN06) to perform fully-parallelized three-dimensional radiation 
transport for a point source in spherical polar grids and for plane waves in cartesian boxes.  Our calculations 
are intended to survey the qualitative nature of three-dimensional instabilities in galactic ionization fronts 
rather than provide detailed comparisons to analytical studies, which have only been performed in one dimension.  
We focus on dynamical phenomena in r$^{-2}$ density fields for comparison to earlier two-dimensional numerical 
work and for their similarity to clump profiles observed in massive star forming regions within molecular clouds 
in the galaxy today \citep{ag85,het88}.  Our numerical suite is therefore of direct relevance to the environments 
of massive stars as well as the morphologies of emission nebulae in the interstellar medium (ISM).  Since our 
simulations all employ radiative cooling in uniform or r$^{-2}$ densities, we expect initial perturbation growth
to be dominated by Vishniac modes.  In nearly adiabatic primordial gas with steeper gradients such as those in
cosmological minihalos, unstable modes could be initiated through the Rayleigh-Taylor channel, a possibility we
investigate in a following paper.

In Section 2 we describe the improvements to our radiative transfer and reaction network integration scheme  
required by transport in three dimensions as well as the parallelization and load balancing of the transport 
itself.  In section 3 we compare two-dimensional calculations of dynamical instabilities on polar grids 
performed with simplified ionization equilibrium physics to those with full radiative transfer.  Our 
three-dimensional simulations of these instabilities on spherical-polar coordinate grids are detailed in 
Section 4 and we present three-dimensional calculations of the shadow instability \citet{rjw99} in section 5.
Implications of this work are discussed in Section 6.

\section{ZEUS-MP Algorithm Upgrades}

The ZEUS-MP hydrocode utilized in previous one-dimensional work \citep{wan04} is now fully parallelized 
for three-dimensional applications.  We implemented the transport of UV plane waves along the x-axis of a
cartesian box (or along the z-axis of a cylindrical coordinate grid) in addition to the original radial 
transfer of photons from a point source centered in a spherical polar coordinate mesh.  The parallel 
reactive flow, additions to the radiative transfer, and improvements to our subcycling scheme are detailed
below.

\subsection{Parallelization Scheme}

We refer the reader to WN06 for a description of our sequential multistep radiation algorithm.
ZEUS-MP decomposes the simulation domain into subunits called tiles, with each tile assigned to a single 
processor for the duration of a calculation \citep{jch06}.  Global solution concurrency is maintained
by message-passing between processors through calls to the Message Passing Interface (MPI) library.  In 
general, the computational domain can be partitioned along all three axes but in photoionization calculations
we limit the decomposition to two axes only. In cartesian coordinates we subdivide the volume along the y and 
z directions only; in cylindrical coordinates the decomposition is along the r and phi axes, and in spherical 
grids we tile only along theta and phi.  Hence, a tile in a cartesian grid is a rectangular box whose length 
in the x direction spans the entire grid.  In cylindrical coordinates, tiles are wedges in r and phi lengthwise 
in z, and in spherical coordinates they are slices in theta and phi radiating outward from the coordinate center 
to the outer radial boundary.

Domain decomposition is implemented along only two coordinates in ZEUS-MP to enforce load balancing of the 
radiative transfer and to avoid the need to communicate photon fluxes between adjacent tiles.  The key to
effective load balancing is requiring that the ionization front be present in all the tiles at once.  
Subdivision along all three axes for a single point source or plane wave would result in the few tiles 
hosting the front shouldering a disproportionate fraction of the chemistry subcycling, leaving the other tiles 
relatively idle.  The tradeoff in this scheme is the limit to the parallelization that can be applied
to a given problem.

The reaction network is evolved plane by plane within a tile.  Updates to the reaction network can be made 
one plane of zones at a time because the chemistry in any given zone over $\Delta$t$_{chem}$ does not depend on its 
nearest neighbors.  We advance the network one plane at a time so the reaction coefficients can be 
stored in two-dimensional arrays.  Up to thirty rate coefficients are present in the reaction network and 
as many as 21 heating and cooling coefficients comprise the isochoric update to the gas energy.  Any network 
update order requiring the coefficient arrays to be three-dimensional would inflate the memory required for 
a calculation by a factor of ten beyond what is required just for hydrodynamics.  

We prevent the solution for the entire grid from being evolved further than the shortest evolution time in 
any of its tiles by performing an MPI all\_reduce operation to extract the global minimum of $\Delta$t$_{heat}$.
The smaller of the grid minima of $\Delta$t$_{heat}$ and the Courant time is adopted for the hydrodynamical
update of gas energies, densities, and velocities.  To further preserve solution coherence, mpi\_barrier
calls are stationed at the end of the reaction network module to ensure subcycling is complete in all the 
tiles before any of them advance to the next operator-split step in the hydrodynamics.  This procedure is
necessary because of the implementation of asynchronous message passing in ZEUS-MP, in which code execution
in some tiles can proceed while communication between other tiles is still in progress.

\subsection{Radiative Transfer: Cartesian and Cylindrical Coordinates}

We compute the attenuation of the ionizing flux along rays parallel to the grid lines.  In cartesian 
coordinates, the rays are parallel to the x-axis.  In cylindrical coordinates, the rays are parallel 
to the z-axis.  Generalizing the radiative rate coefficients k$_{rad}$ evaluated in WN06 to 
planar radiation flows is straightforward.  Recalling that the number of photons removed from a zone 
per second (n$_{ioniz}$) is related to the radiative rate coefficient (k$_{rad}$) by eq. 18 of WN06
\begin{equation}
k_{rad} = \displaystyle \frac{n_{ioniz}}{n_{H}\,V_{cell}},
\end{equation}
and that photon conservation demands that \vspace{0.1in}
\begin{equation}
n_{ioniz} = \displaystyle{\frac{F_{out}A_{out} - F_{in}A_{in}}{h\nu}},\vspace{0.1in}
\end{equation}
where F and A are the flux (in erg/sec/cm$^2$) and area over the given face of the zone, we have that \vspace{0.1in}
\begin{equation}
k_{rad} = \displaystyle{\frac{{\dot{n}}_{ph} \left( 1 - e^{-\chi \left(r_{i+1}-r_{i}\right)}\right)}{n_H \Delta x}},\vspace{0.1in}
\end{equation}
where ${\dot{n}}_{ph}$ is the emission rate of all photons at a given energy, $\Delta$x is the zone 
length along the x-axis, and $\chi$ is the mean free path of the photons in the neutral gas.  This 
derivation holds with axial plane wave transport in cylindrical geometries because the metric terms in 
the area factors drop out.  We include an option for attenuating the plane wave flux by r$^{-2}$ to
approximate the geometric dilution of a point source.

\begin{figure}
\resizebox{3.45in}{!}{\includegraphics{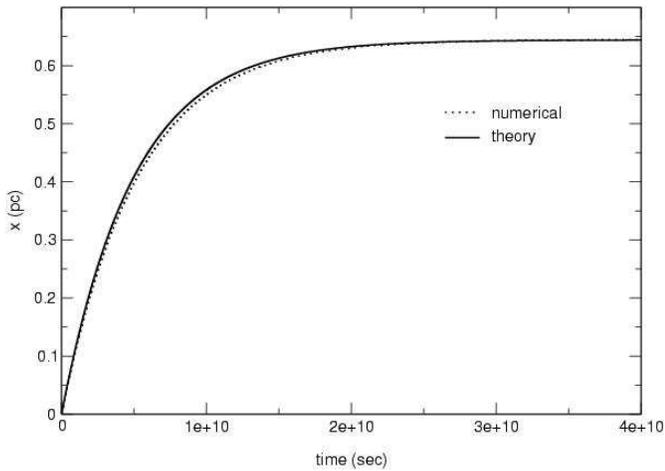}}
\caption{Approach of the static planar ionization front to x$_{Str}$, numerical vs analytical.} \vspace{0.25in}
\label{fig:zstrom}
\end{figure}

We tested the radiative transfer by computing the position of a one-dimensional planar ionization front 
in a uniform static medium of hydrogen, with n$_H$ = 1000 cm$^{-3}$ and T = 72 K.  The grid has 1000 
zones in the x direction and a length of 1.0 pc.  The photon flux along the x-axis incident to the 
left-hand face is J$_0$ = 4.0 $\times$ 10$^{11}$ cm$^{-2}$ s$^{-1}$, and we applied a constant case B 
recombination coefficient $\alpha_B$ = 2.015 $\times$ 10$^{-13}$ cm$^3$ s$^{-1}$.  By equating the photon 
flux at the front to the incoming flux of neutral particles we can compute the velocity and position of 
the front in the usual manner:\vspace{0.1in}
\begin{equation}
v_{f} \; = \; \displaystyle\frac{m_{i}}{\rho_{i}}\;J_{f} \; = \; \displaystyle\frac{n_{\gamma,f}}{n A} = \; 
\displaystyle\frac{1}{n} \; \left[J_0 - x_{f} n^2 {\alpha}_B \right] \vspace{0.1in}
\end{equation}
where n$_{\gamma,f}$ is the total number of ionizing photons reaching the front per unit time and A is 
the area of the zone face.  The position of the front easily follows from simple integration:\vspace{0.07in}
\begin{equation}
x_f \; = \; \displaystyle\frac{J_0}{n^2 {\alpha}_B} \left( 1 - e^{-\alpha_B t} \right) \vspace{0.07in}
\end{equation}
We show the evolution of the front in the static medium in Fig \ref{fig:zstrom}; for the parameters of 
this test, x$_{Str}$ = 0.644 pc.  The code agrees with the analytical prediction to within 5\% at all
times.  A parallelized version of this test was performed in three dimensions to verify that the advance 
of the front in each tile was identical.

\subsection{Advanced Subcycling}

As detailed in WN06, in one-dimensional problems the primordial gas reaction network in ZEUS-MP 
was cycled over consecutive chemical timesteps \vspace{0.05in}
\begin{equation}
\Delta t_{chem} = 0.1 \, \displaystyle\frac{n_{e}}{{\dot{n}}_{e}} \vspace{0.05in}
\end{equation}
until a hydro timestep was covered \vspace{0.1in}
\begin{equation}
\Delta t_{hydro} = min(\Delta t_{cour}, \Delta t_{heat}) = min \left(\displaystyle\frac{\Delta\,r}{c_{s}},\, 
0.1 \displaystyle\frac{e_{gas}}{{\dot{e}}_{ht/cl}} \right) \vspace{0.1in}
\end{equation}
This scheme proved to be robust and versatile, accommodating regions experiencing either 
photoionization or recombination.  However, timescale analysis revealed that cells close to the source
without any initial electron fraction required as many as 70 - 80 hydrodynamical updates before 
coming to ionization equilibrium upon arrival of the front.  Network subcycling between updates also 
tended to be extreme in such zones, with several thousand iterations required to cross $\Delta$t$_{hydro}$ at
times, far more than are required for accuracy.  The problem is ameliorated at larger radii because 
heating times increase further away from the source.  Also, advection transports small electron 
fractions into zones slightly ahead of the front, preventing initially minute $\Delta$t$_{chem}$.

Practical I-front transport in three dimensions required two improvements to our integration scheme.
When the front encounters overdensities, advection of electrons into those zones is halted, resulting 
in paralyzingly short chemical timesteps until the front exits the perturbation.  We eliminated this
problem by modifying $\Delta$t$_{chem}$: \vspace{0.05in}
\begin{equation}
\Delta t_{chem} = 0.1 \, \displaystyle\frac{n_{e} + 0.001 n_{H}}{{\dot{n}}_{e}} \vspace{0.05in}
\end{equation}
The correction term ensures that $\Delta$t$_{chem}$ is not too small when photoionizations commence in the zone
and has little effect as the zone approaches equilibrium and $\Delta$t$_{chem}$ increases.  The algorithm now 
advances ionization fronts through density jumps without difficulty.

We also experimented with more agressive subcycling by requiring that a zone come to ionization 
equilibrium in no more than 7 - 8 hydrodynamical updates by altering $\Delta$t$_{hydro}$: \vspace{0.1in}
\begin{equation}
\Delta t_{hydro} = min\left(\Delta t_{cour}, \displaystyle\frac{e_{gas}}{{\dot{e}}_{ht/cl}} \right). \vspace{0.1in}
\label{eqn: thydro}
\end{equation}
We obtain speedups of up to 15 in transporting R-type fronts with this time step prescription, but 
there is little improvement with D-type fronts because in primordial environments $\Delta$t$_{cour}$ is 
usually the smaller of the two timescales in eq. \ref{eqn: thydro}.  Two difficulties arose in this
method.  First, the longer time step can cause overshoot in the cooling as the zone reaches full
ionization, occasionally removing all the gas energy in a zone in a single update.  We resolved this 
issue by reverting to the original time step if ${\dot{e}}_{gas}$ $<$ 0 in a zone, which removed the
anomalous cooling with little sacrifice in performance.

Unfortunately, the coarser time steps can lead to inaccuracies in the reaction network, causing the
propagation of R-type fronts to diverge from theory by as much as 10 - 20\%.  In many practical 
applications this is not a serious error, so the performance gain may outweigh the loss in 
accuracy.  While we retain both subcycle options in the code, we applied only the improved $\Delta$t$_{chem}$
to these studies to properly capture peturbations in R-type fronts that later grow into instabilities 
after becoming D-type.

\section{D-Type Instabilities in Two Dimensions}

We now examine the growth of unstable modes in a D-type ionization front that is expanding in a radially 
symmetric gas profile randomly seeded with minute fluctuations in density on a polar coordinate grid.  We 
consider the development of instabilities in a uniform core enclosed by an r$^{-2}$ envelope
\[ n_{H}(r) = \left\{ \begin{array}{ll}
			   n_{c}                 & \mbox{if $r \leq r_{c}$} \\
			   n_{c}(r/r_{c})^{-2}   & \mbox{if $r \geq r_{c}$}
                          \end{array}
                  \right.\vspace{0.1in} \] \label{ngas}
assuming that the shocked gas is radiatively cooled by metal emission lines.  This problem was first studied 
by \citet{gsf96} (hereafter GSF) and we compare their results below.

We augmented the isochoric gas energy update in ZEUS-MP (eq. 9 of WN06) with cooling curves from 
\citet{dm72} in which any elemental abundances can be specified.  The Dalgarno \& McCray curves include both
electron and neutral hydrogen collisional excitation of H, He, and neutral and singly-ionized C, O, N, Fe, Si, 
and S.  The energy equation also has cooling due to electron collisional ionization of hydrogen, H and He 
recombinational cooling, bremsstrahlung cooling, and Compton cooling by free electrons off the cosmic microwave 
background (for high-redshift applications).  These latter two processes play a negligible role in our simulations.  
In general, the Dalgarno \& McCray cooling rates in a given cell can be constructed with any electron or neutral 
H number density derived from the reaction network.  For consistency with GSF, we evaluate them assuming a fixed 
electron fraction of 0.01 and solar metallicity in the calculations in this paper, regardless of the true electron 
abundances on the grid.  

This treatment approximates radiative cooling in shocked neutral gas well but underestimates metal line cooling, 
H-$\alpha$ cooling, and He-$\alpha$ cooling in ionized gas because the electron fraction is much higher than 0.01  
there.  At postfront gas temperatures above $\sim$ 10$^4$ K, H-$\alpha$ and He-$\alpha$ cooling rates per atom can 
exceed those of the metals by four orders of magnitude.  Given that neutral H and metal fractions in the ionized 
gas at solar metallicities are roughly equal ($\sim$ 10$^{-5}$), H-$\alpha$ and He-$\alpha$ cooling dominates the 
total Dalgarno \& McCray curve inside the H II region.  However, in the pure H simulations of this study we incur 
overestimates of at most 10\% in temperature in the ionized gas because it is H recombinational cooling that
really controls postfront temperatures, as we demonstrate below.  Cutoff temperatures were also incorporated in 
the runs, defined to be the gas temperature below which cooling is not applied to the gas energy density.  
For the two runs discussed in this section we adopted cutoff and background of temperatures of 100 K.  

A point source of ionizing UV flux was centered on the polar coordinate grid with 200 radial zones and 180 
zones in the theta direction.  The inner and outer radial boundaries were 0.01 pc and 2.0 pc with reflecting 
and outflow boundary conditions, respectively.  The inner and outer boundaries in the theta direction were 
$\pi$/4 and 3$\pi$/4, respectively, with reflecting boundary conditions.  We adopted this angle range to avoid 
the minute Courant times associated with zones near the coordinate poles due to their small angular dimensions.  
At the resolution of this simulation we obtain minimum Courant times ten times greater over this range than at 
the poles.  Our choice of reflecting boundary conditions could exclude the growth of long-wavelength modes, 
but the linear stability analysis of \citet{gu79} predicts that short-wavelength modes will initially dominate 
and then later consolidate into larger structures.  

The density profile in both runs was hydrogen only, with a flat central core of 1.0 $\times$ 10$^4$ cm$^{-3}$ 
out to r = 0.2 pc followed by an r$^{-2}$ dropoff.  At the outer boundary the density falls to $\sim$ 100 cm$^{-3}$.  
The initial gas temperature was set to 100 K.  We adopted a photon emission rate ${\dot{n}}_{ph}$ of 1.0 $\times$ 
10$^{48}$ s$^{-1}$ with energy 15.2 eV, with an energy per ionization $\epsilon_{\Gamma}$ of 1.6 eV in order to 
maintain the ionized gas temperature at 1.0 $\times$ 10$^4$ K.  Theory predicts that if the initial Str\"{o}mgren 
radius falls in the density gradient the I-front will revert from D-type to R-type and overrun the envelope 
\citep{ftb90}.  For the central density and emission rate of these runs, the Str\"{o}mgren radius  
\begin{equation}
R_S = \left(\displaystyle\frac{3 {\dot{n}}_{ph}}{4 \pi {\alpha}_B {n_{c}^2}} \right)^{1/3}\vspace{0.1in}
\end{equation}
is 0.065 pc, entirely within the central core.  The transition to R-type is thus delayed.

In the first run we do not perform radiative transfer but instead repeat the simplified treatment applied in
the GSF models.  Every hydrodynamical time step the equilibrium position of the I-front is computed along a 
given radial ray by equating the sum of the case B recombinations along the ray to the emission rate of the 
central source:\vspace{0.1in}
\begin{equation}
\int_{0}^{r_f} 4 \pi r^{2} n_{e} n_{p} {\alpha}_B dr = {\dot{n}}_{ph}\vspace{0.1in}
\end{equation}
The gas temperature is then set to 1.0 $\times$ 10$^4$ K along the line of sight out to r$_f$.  After repeating
this procedure for every value of theta the model is evolved for a Courant time.  The position of the front
thus changes each time step as gas is rearranged on the grid by the high-temperature flow.  We refer to this 
approach in the following as ``ionization equilibrium''.  In practice, we 
found it necessary to limit the cooling in a shocked zone to be no greater than 90\% of the energy remaining 
in the zone above the temperature cutoff of the cooling curve.  Otherwise, the rather coarse hydrodynamical 
timesteps can allow explicit updates to the energy equation to completely remove all energy within the cell.
No R-type fronts ever arise in this approximation since the fronts by construction are always assumed to have 
reached the Str\"{o}mgren at any given time step.

In the second run we transport the front across the grid using radiative transfer scheme described in Sec. 2.2,
coupled to self-consistent photoionization and heating as described in Sec. 2.3 (see also WN06).  In the \citet{gsf96}
simulations, density fluctuations with randomly-distributed amplitudes of at most 1\% were imposed on very 
zone to seed the formation of shock instabilities.  We also did this, holding the gas energy density constant 
to prevent pressure fluctuations from dispersing the perturbations.  However, in our model we only imposed 
these variations beyond radii of 0.125 pc to prevent the onset of shadowing instabilities in the early R-type 
front before dynamical instabilities in the D-type front can form.  We compare the evolution of both models
in Fig \ref{fig:2D}.

\subsection{Ionization Equilibrium vs Radiation Hydrodynamics}

\begin{figure*}
\epsscale{0.8}
\plotone{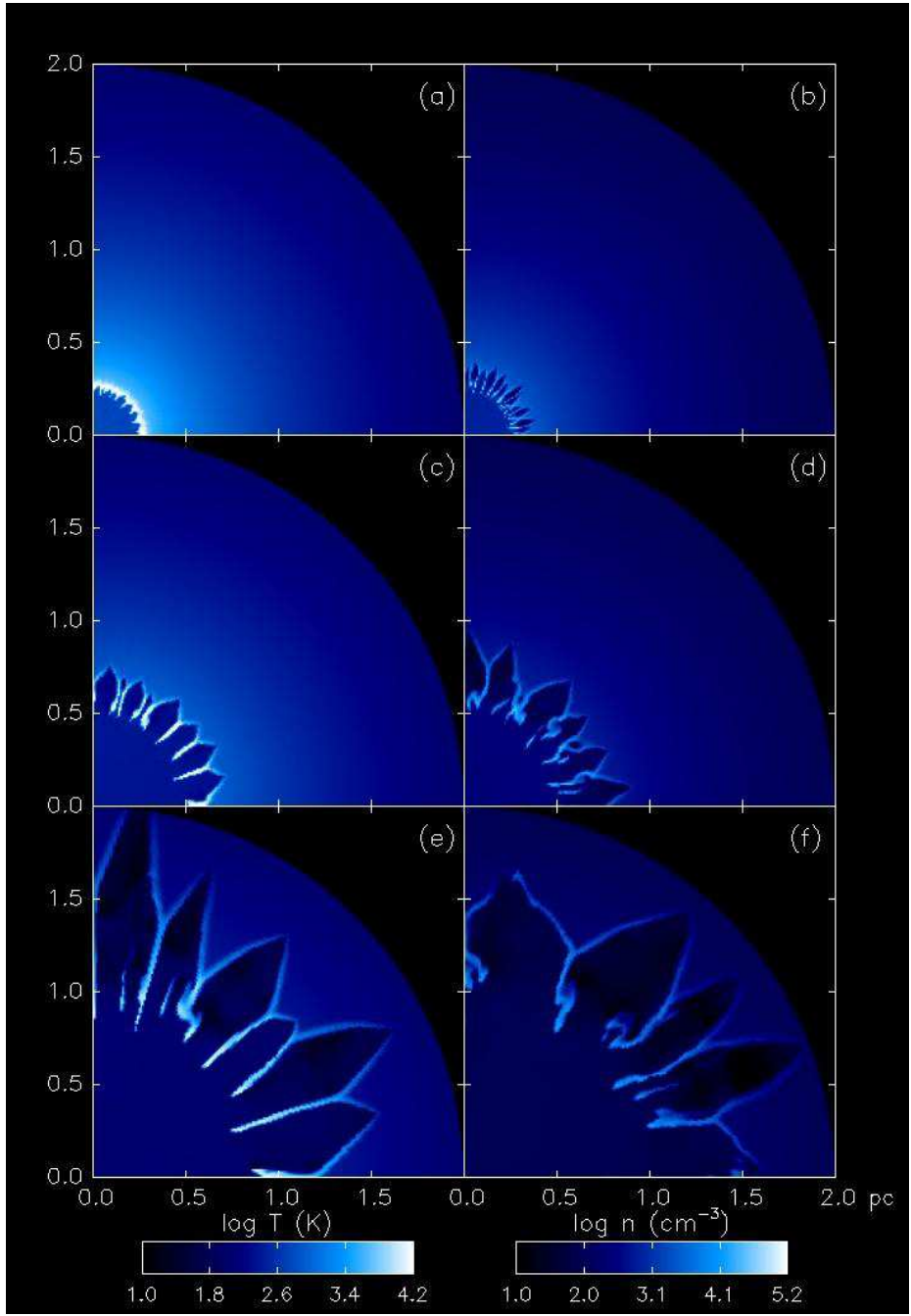}\vspace{0.15in}
\caption{Dynamical instability in a D-type I-front propagating out of an r$^{-2}$ density gradient with
metal line cooling.  Panels (a), (c), and (e): number densities at 28.5 kyr, 76.1 kyr, and 148.4 kyr 
for a front evolved assuming simple ionization equilibrium.  Panels (b), (d), and (f): density evolution 
of an I-front transported across the grid with full radiative transfer at 28.5 kyr, 76.1 kyr, and 148.4 
kyr. 
\label{fig:2D}} 
\vspace{0.075in}
\end{figure*}

Both models confirm the prediction of \citet{gu79} that the modes which grow fastest are those 
with the largest wavelengths.  Unstable modes begin to disrupt the shock at 9.5 kyr in the second run, 
with 13 spikes visible at 28.5 kyr (panel (b)) but only 5 remaining at 148.4 kyr (panel(f)).  Vorticity 
in the flow is evident where the bases of the spikes join the flared regions.  The instabilities remain 
D-type:  they are always bounded by a dense shocked gas layer and their ionized interiors have time to 
laterally expand as they elongate.  The perturbation amplitudes never saturate in either model but 
continue to grow well into the nonlinear phase, from 0.1 pc at 28.5 kyr to 0.5 - 1.0 pc at 148.4 kyr in 
the second run.  Direct quantification of their growth rates is problematic because individual structures 
coalesce over time, but their rapid elongation at later times is due to their descent down the density 
gradient. 

There are noticeable differences in structure of the fronts at early and intermediate times.  While the 
ionized fingers are numerous and narrow in both simulations at early times (panels (a) and (b) of Fig 
\ref{fig:2D}), their morphologies diverge after a few tens of kyr.  Ionized spikes push past each other 
at random times through the shock in the second run, flaring sideways earlier than in the first run 
because they are not as collimated by their neighbors.  In contrast, the spikes advance nearly in step 
with one another in the first model, remaining cylindrical out to much greater distances because of the 
parallel advance of their neighbors.  There are also less of them: 7 spikes at 0.3 pc in comparison to 
13 in the second simulation.  The differences between the two models are most pronounced at early and 
intermediate times, becoming less distinct as they enter highly nonlinear regimes.  
  
\begin{figure}
\resizebox{3.45in}{!}{\includegraphics{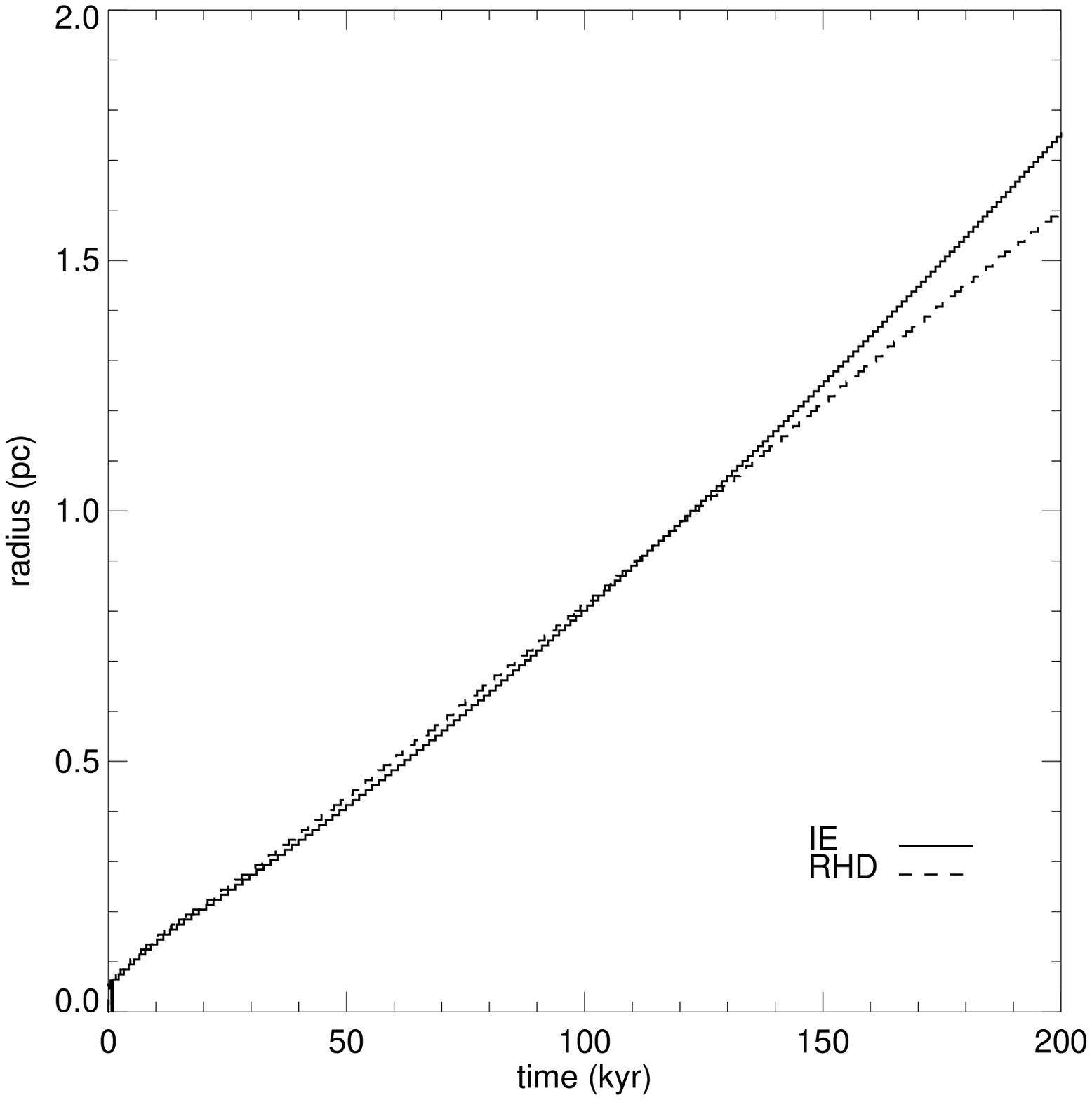}}
\caption{One-dimensional I-front positions in the ionization equilibrium approximation and with
radiative transfer} \vspace{0.25in}
\label{fig:2Dp}
\end{figure}

The departures are chiefly due to how the two numerical schemes capture the interaction of the front 
with the shocked neutral shell.  The topology of the instability is sensitive to how the inner shell 
is photoevaporated.  Ionization equilibrium abruptly assigns temperatures of 1 $\times$ 10$^4$ K to 
the fluid elements on the inner surface that are capable of being ionized.  Had they instead been 
incrementally heated and ablated, as in the radiation hydrodynamical scheme, their subsequent expansion 
would have created a larger number of cracks in the shell distributed over time, yielding the larger 
number of ionized fingers observed in the second run.  Although costlier, full radiation transport 
is necessary to model the actual deformation of the shocked shell.  Initial disparities in the 
deformation of the shell by the two algorithms lead to the much greater differences in morphology at 
intermediate times evident in panels (c) and (d) of Fig \ref{fig:2D}.  

In one-dimensional calculations the two algorithms predict I-front positions to within a few percent
of each other at early times but diverge by more than 10\% at later times, as shown in Fig \ref{fig:2Dp}.  
This agreement holds only so long as the front is D-type.  Ionization equilibrium approximations 
would not fare as well if the front became D-type in the r$^{-2}$ gradient or in other stratified 
media.  It would fail altogether if the front executes transitions from D-type to R-type or vice 
versa, as is common in cosmological H II regions.  We have found that updates to velocities should 
be performed on photoheating time scales rather than the Courant time to capture the true acceleration 
of fluid elements by the front in power-law density gradients (WN06).

\section{Three-Dimensional D-Type Instabilities}

\begin{figure*}
\epsscale{0.8}
\plotone{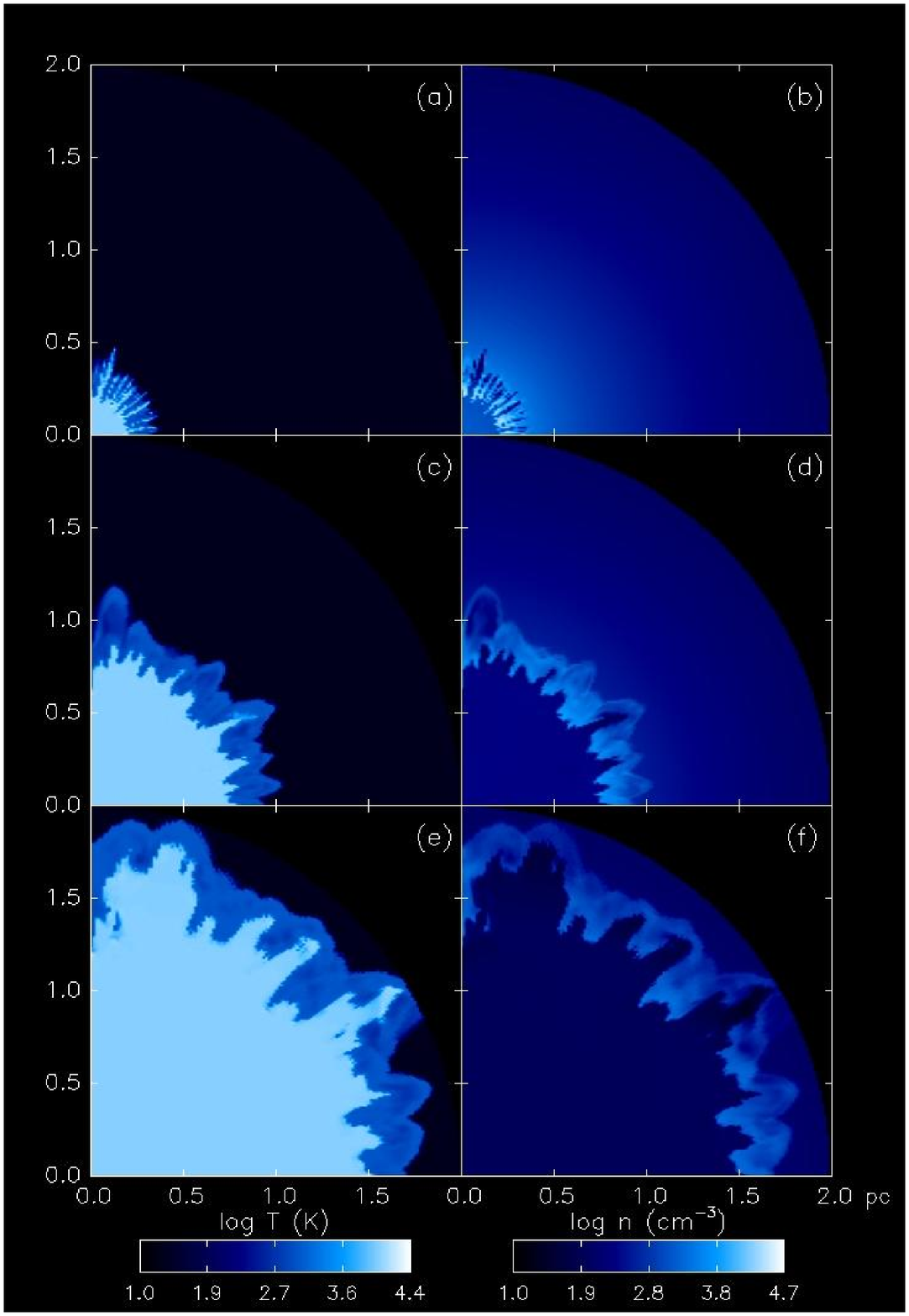}\vspace{0.15in}
\caption{Dynamical instability in a D-type front propagating outward along an r$^{-2}$ density gradient with
metal line cooling, run S31.  Panels (a), (c), and (e): temperature evolution at 20.8 kyr, 96.2 kyr, and 177.4 
kyr.  Panels (b), (d), and (f): density evolution at 20.8 kyr, 96.2 kyr, and 177.4.\label{fig:S31DM}} 
\vspace{0.075in}
\end{figure*}

\begin{figure*}
\epsscale{0.8}
\plotone{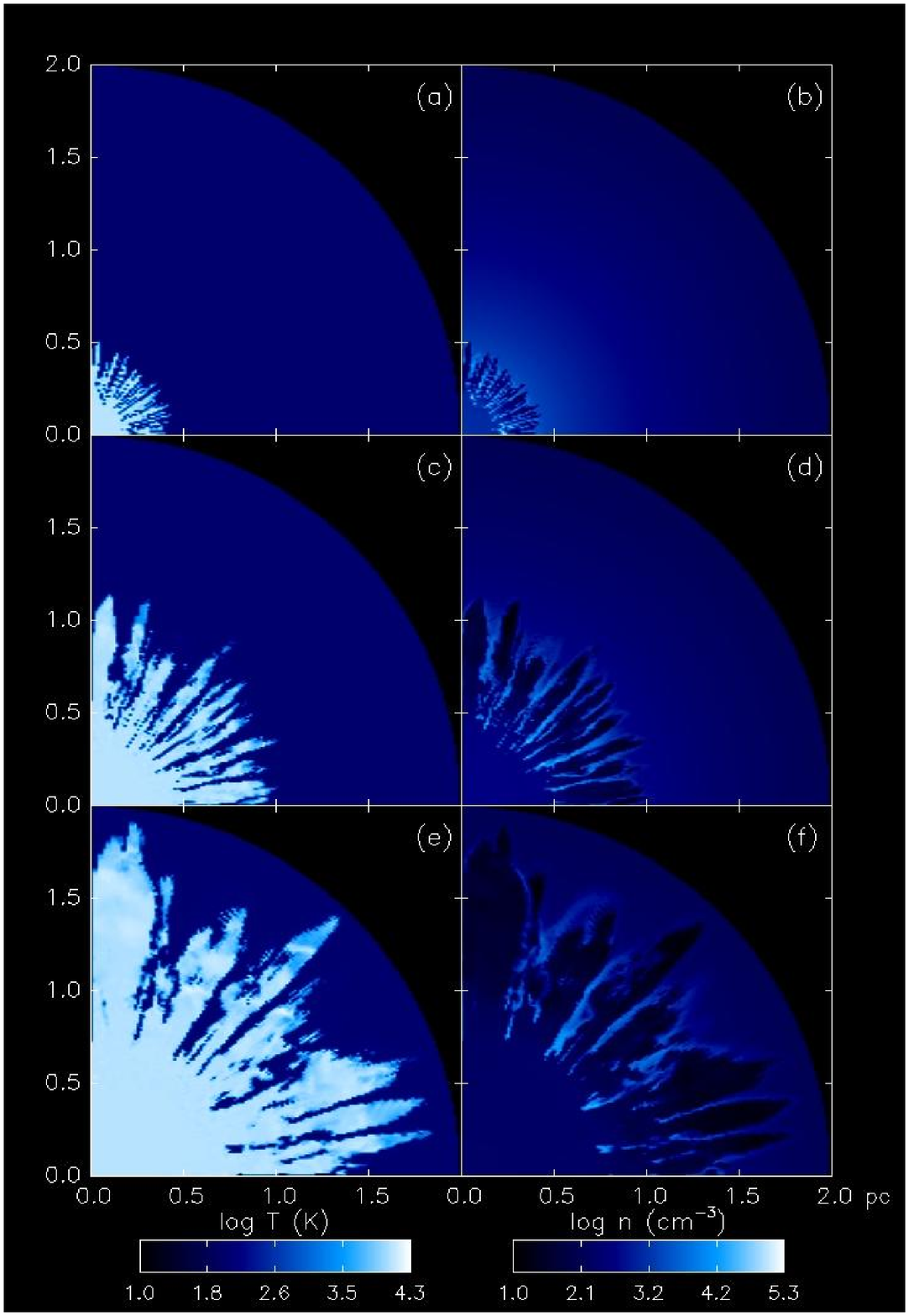}\vspace{0.15in}
\caption{D-type ionization front instability in an r$^{-2}$ density gradient with metal line cooling, run S22.
Panels (a), (c), and (e):  temperature evolution at 20.8 kyr, 67.9 kyr, and 124.5 kyr.  Panels (b), (d), and 
(f):  density evolution at 20.8 kyr, 67.9 kyr, and 124.5 kyr.\label{fig:S22}} 
\vspace{0.075in}
\end{figure*}

We extended the radiative transfer calculation in section 3 to three dimensions using a spherical 
polar coordinate grid with 200 radial zones and 180 zones in the theta and phi directions.  The inner and 
outer radial boundaries were again 0.01 pc and 2.0 pc with reflecting and outflow boundary conditions, 
respectively.  The angle boundaries were again chosen to be $\pi$/4 and 3$\pi$/4 in phi and theta with 
reflecting boundary conditions.  We apply the same randomly perturbed density profile as before to the 
grid, with the energy per ionization $\epsilon_{\Gamma}$ set to 1.6 eV to ensure a postfront gas temperature 
of 1.0 $\times$ 10$^4$ K.

Our first run (S31) had cutoff and background temperatures of 1000 K and 10 K, respectively, while our 
second run (S22) had cutoff and background temperatures of 100 K.  Metal lines cannot fully cool shocked 
gas in the first case so the shell is thinner than in an adiabatic flow but not as dense (or cold) as when 
the metals are free to cool the gas down to the background temperature.  \citet{gsf96} nevertheless found 
instabilities to readily form in these circumstances in their two-dimensional simulations.  The lower cutoff 
in the second run is intended to produce a relatively cold, dense shell that will break up more violently. 
In both of these models ${\dot{n}}_{ph}$ was again chosen to be 1.0 $\times$ 10$^{48}$ s$^{-1}$ so that the 
Str\"{o}mgren sphere would reside within the core.

\subsection{S31 Instability}

In Fig (\ref{fig:S31DM}) we show $\phi$ = $\pi$/2 slices of the evolution of the ionization front in the S31 
run at 29.5 kyr, 70 kyr, and 100 kyr (the total time for which the front was followed in the 1996 experiments).  
In panels (a) and (b) the perturbations are seen to be dominated by short-wavelength modes at early times that 
later merge into the longer-wavelength modes in panels (c) and (d), in accord with linear stability analysis.  
The initial wavelengths are of the order of the shell thickness.  Unable to completely cool, the shocked shell 
becomes thicker as neutral gas accumulates onto it, acquiring turbulent motions as it evolves.  The instabilities 
persist to 200 kyr in these three-dimensional calculations as seen in panels (e) and (f) but saturate without 
the prominent growth exhibited in the GSF two-dimensional models.  The turbulence in the shell appears to 
prevent the unstable modes from breaking out.  The maximum density attained by the clumps in the course of the 
simulation is 4.99 $\times$ 10$^4$ cm$^{-3}$; pronounced clumping is visible at 177 kyr in panel (f). 
  
\subsection{S22 Instability}

The ionization front breaks through the shock and escapes much further down the stratified envelope in the 
S22 run, as dramatically illustrated in Fig (\ref{fig:S22}).  We show $\phi$ = $\pi$/2 slices of the front 
for this model at 20.8 kyr, 67.9 kyr, and 124.5 kyr.  Radiation flares outward through the cracks in this 
more violently fragmenting, colder, denser shell, and the dominant growth modes over time are again those 
with the greatest wavelengths.  This motion is visible in the large dark neutral patches cutting across 
some of the ionized flares in temperature panel (e) of Fig (\ref{fig:S22}).  Comparison of panels (c) and 
(e) demonstrate the transience of these features.  Newly ionized gas in the outer regions of the flares 
expand laterally as the instabilities propagate, creating dark, rarefied zones in the corresponding density 
panels.  The unstable modes advance as D-type, given their time scales of expansion as well as because they 
remain bounded by shocked gas that is visible as a thin, lighter hued density layer in the outer edges of 
the flares.  The maximum density of the gas in this run is 2.08 $\times$ 10$^5$ cm$^{-3}$, greater than in 
the previous run due to the higher degree of postshock cooling.  The thin radial segments of cool zones
visible in temperature panels (c) and (e) are cold, dense shell fragments of the shocked shell not yet 
ionized by the front.  These fragments can be seen to originate in the cold shell because they extend from
the base of the instabilities and radiate outward.  They are overtaken on either side by hot ionized 
outflows, with the neutral parcels of gas closest to the central source eventually becoming ionized themselves.
Their large densities stand out in relief against the surrounding dimmer regions in panels (d) and (f).  
The cooling of the fragments down to the cutoff temperature is due their high densities rather than some 
instability in the explicit updates to the energy equation.  Our integration scheme prevents any energy 
loss greater than 10\% over the time step on which hydrodynamic updates are performed.  Large clumps of
gas can be seen to persist to late times as they are blown outward by the front in panel (f). 

\begin{figure*}
\epsscale{1.15}
\plottwo{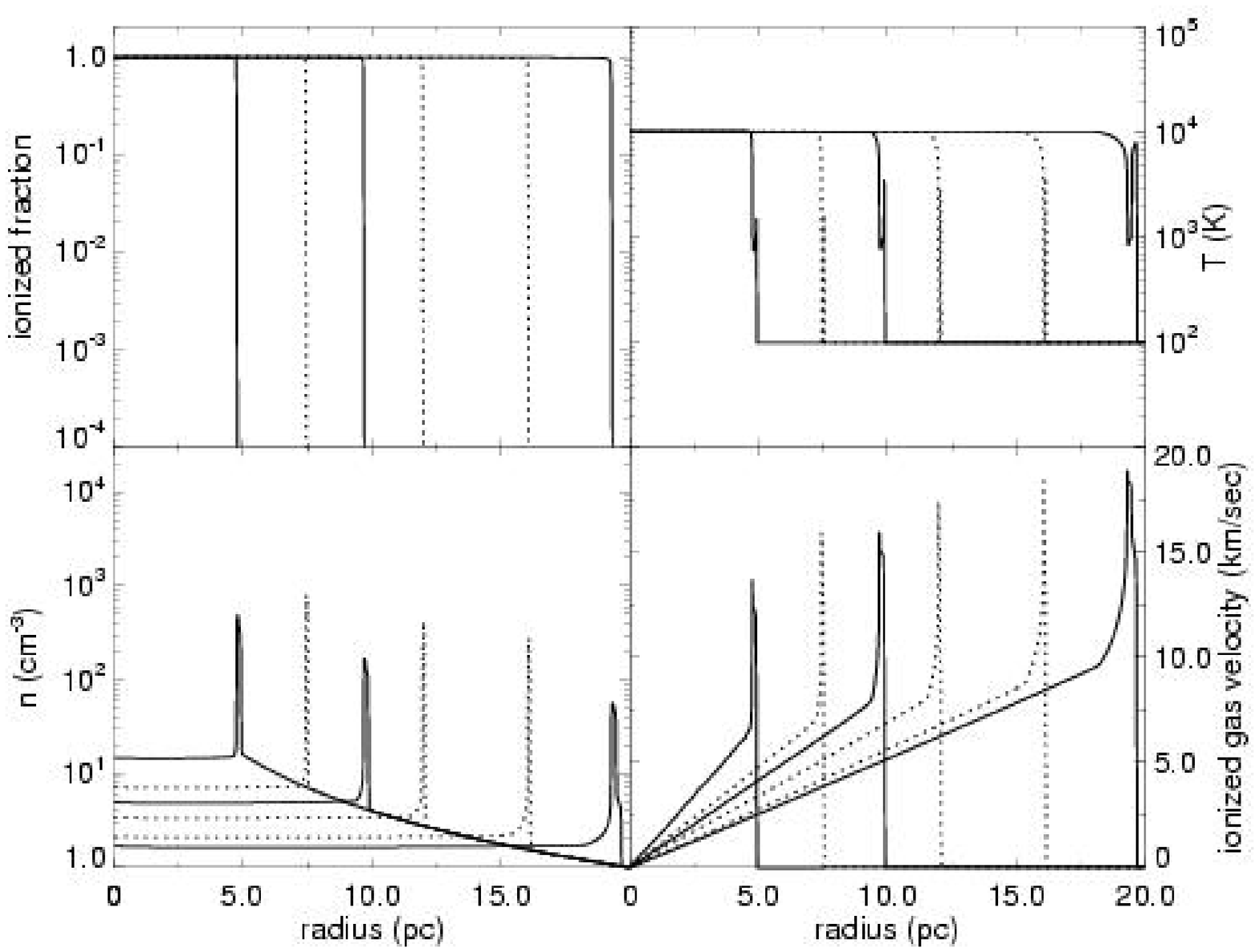}{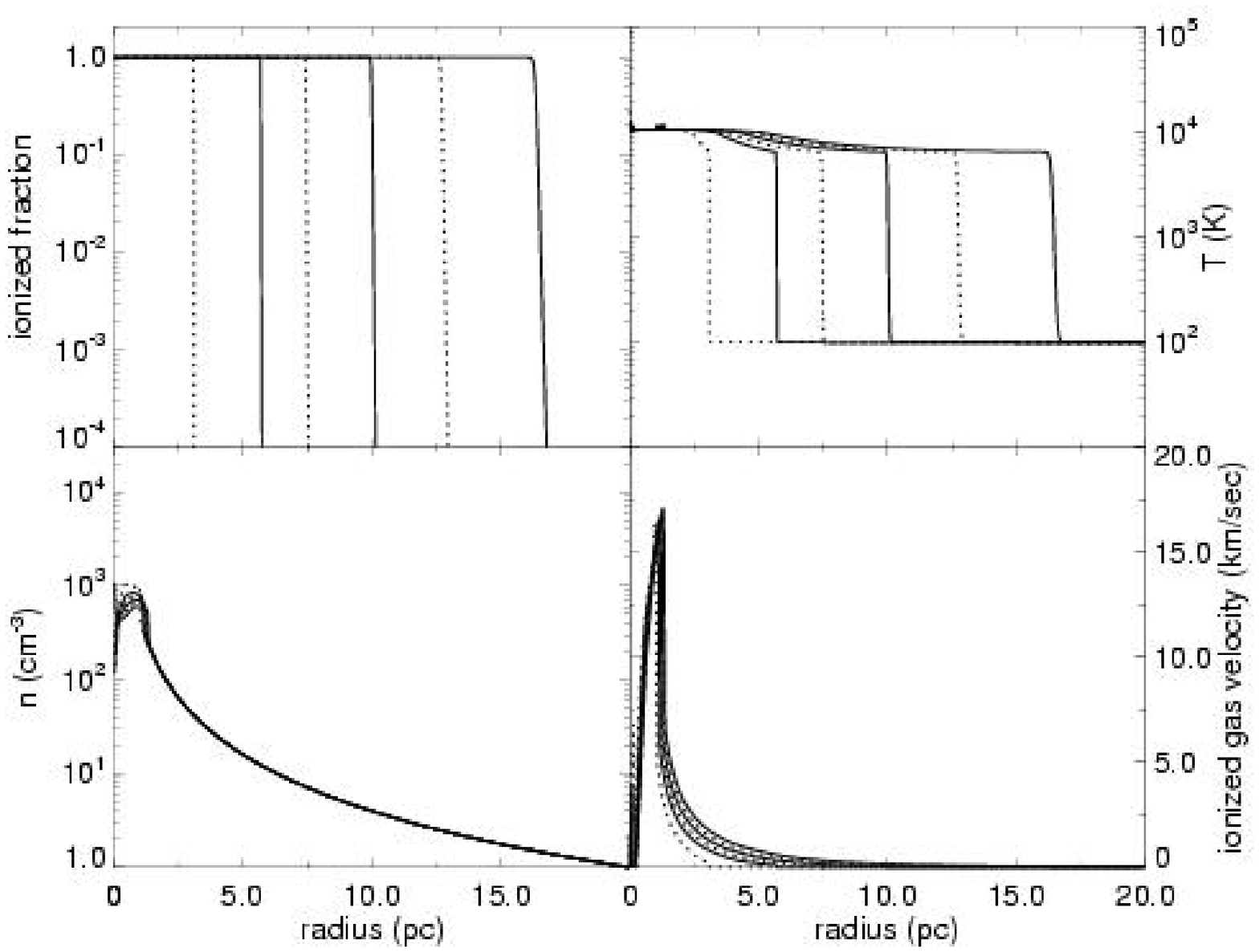}
\caption{Hydrodynamical profiles for ionization fronts in a spherically-symmetric r$^{-2}$ density stratification.
Solid lines are I-front profiles with gas cooling truncated at 1000 K while dotted lines denote profiles in which
gas cooling is halted below 100 K.  Left panel:  R$_S$ $<$ r$_c$.  1000 K profiles are at 0.48 Myr, 0.82 Myr, and 
1.37 Myr while the 100 K profiles are at 0.7 Myr, 1.01 Myr, and 1.26 Myr.  Right panel:  R$_S$ $>$ r$_c$.  1000 K 
profiles are at 36 kyr, 40 kyr, and 44 Kyr while the 100 K profiles are at 32 kyr, 38 kyr, and 42 kyr. 
\label{fig:r2}}  
\vspace{0.1in}
\end{figure*}
\nopagebreak
 
There are clear qualitative differences between the unstable flow in the two-dimensional radiation hydrodynamical model
in the last section and this run that can only be attributed to their dimensionality, given that both
otherwise employ the same physics.  First, transverse flow of gas into the plane of the image from 
above and below is evident in some of the dark patches in the outer regions of the ionized fingers in 
panels (c) and (e).  These are not cold dense shell fragments (as the density panels attest) but instead
are neutral gas crowded into the plane of the image by adjacent instabilities.  Second, the shapes of the 
fingers differ in the two runs: the structures in the three-dimensional model are both more narrow and
plentiful than in the two-dimensional run.  Less flaring occurs in the three-dimensional fingers because
they are partially collimated by neighboring flows that are not possible in the two-dimensional models.  
The geometry of cracks in the shell at the base of the instabilities may also be altered by the additional 
degree of freedom in the flow.

\begin{figure*}
\epsscale{0.8}
\plotone{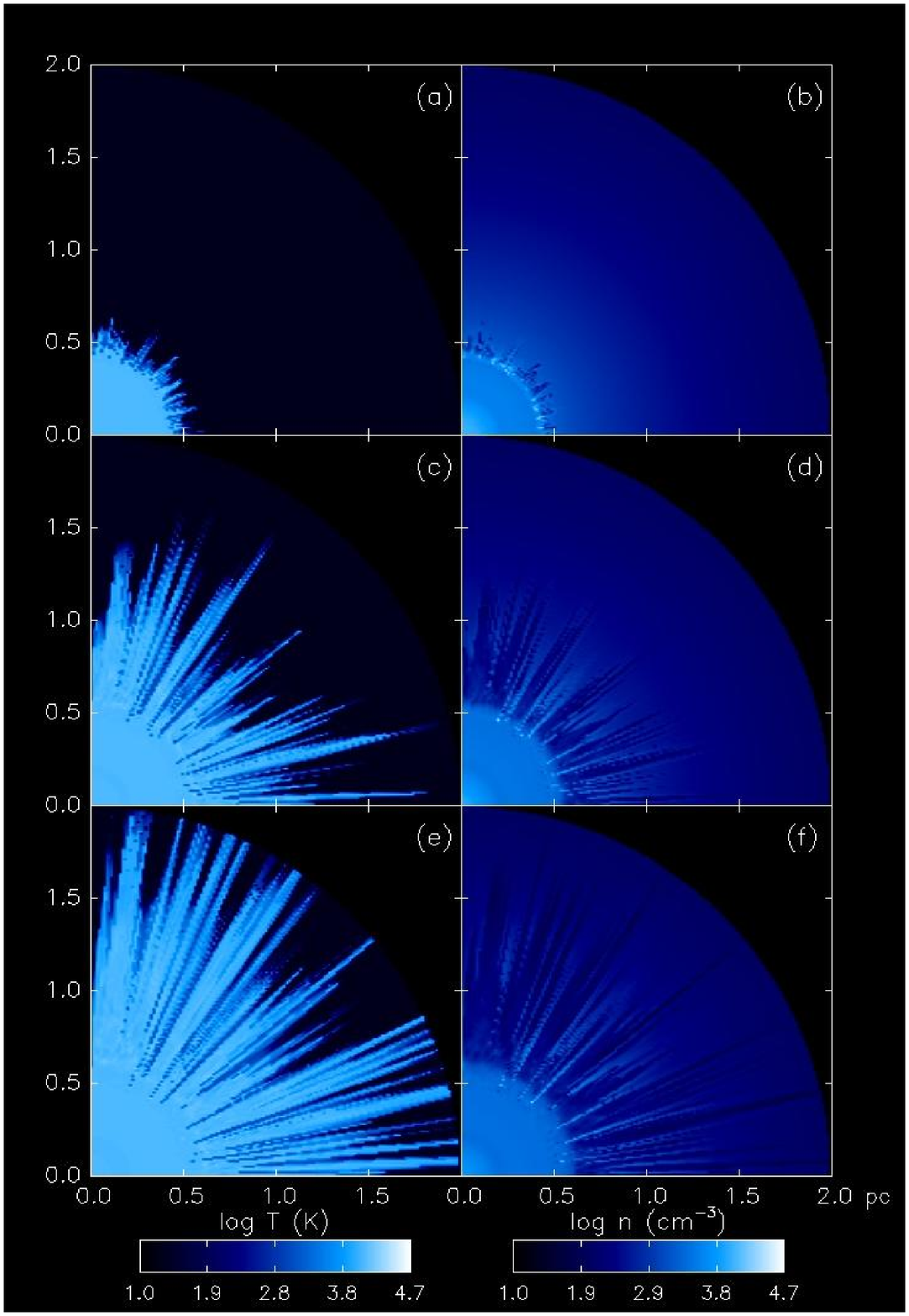}\vspace{0.15in}
\caption{Run S31 with R$_\omega$ $>$ r$_c$.
Panels (a), (c), and (e):  temperature evolution at 9.4, 13.2, and 15.1 kyr.  Panels (b), (d), and (f):  
density evolution at 9.4, 13.2, and 15.1 kyr.\label{fig:S31brk}} 
\vspace{0.075in}
\end{figure*}

The S31 and S22 data confirm that I-front breakout from the envelope is determined by both the degree of 
fragmentation of the shell (which is governed by the efficiency of cooling in the shocked gas) and the 
pressure ratio of the ionized and background gas.  At early stages of the instability there are clear 
differences in the initial scales of the perturbations that we suspect, but have not confirmed, are due 
the thickness of the shocked gas shell at the time of initial fragmentation.  The shell thicknesses vary
from five to ten zones at early times, within the numerical resolution of our mesh.  The initial scales 
of these deformities in part differentiate the later topologies of the S31 and S22 instabilities.

We further note that the absence of runaway instabilities  
in our S31 model may in part be due to R$_S$ being less than r$_c$.  In one dimension, somewhat surprisingly
we find that the density envelope actually confines the front behind the shock at distances far beyond the 
problem boundaries in our calculations, in contrast to the relatively quick breakout that would occur had 
the initial Str\"{o}mgren radius fallen within the gradient.  It is therefore likely that the perturbed 
density profile would prevent further growth of the instabilities until they approach the breakout radii 
found in our one-dimensional calculations discussed below.  Accurate models of I-front transitions are 
therefore key to capturing the correct growth of instabilities and demand full radiation hydrodynamical 
treatment of the ionized flows.  

If the Str\"{o}mgren radius lies beyond the flat central core in the one-dimensional density profile of eq. 
\ref{ngas} then its value is given by \vspace{0.10in}
\begin{equation}
R_{\omega} = R_{S}\left[\frac{3-2\omega}{3} + \frac{2\omega}{3}\left({\frac{r_{c}}{R_{S}}}
\right)^{3}\right]^{1/(3-2\omega)}\left(\frac{R_{S}}{r_{c}}\right)^{2\omega/(3-2\omega)},\vspace{0.10in}
\end{equation}
where R$_S$ is the Str\"{o}mgren radius for the front in a uniform medium.  If R$_{\omega}$ $>$ r$_c$ 
the I-front executes the classical transition to D-type but then soon reverts to R-type as it descends 
the gradient.  The \citet{ftb90} analyses do not address the evolution of the front when it is initially 
confined to the core (R$_S$ $<$ r$_c$), as in these models.  To determine the outcome of the front in 
this regime we performed two sets of one-dimensional calculations in the density profile used earlier 
but without perturbations.  We extended the outer boundary of the mesh to 20 pc with 2000 radial zones 
to preserve the spatial resolution.  One set had a central photon rate ${\dot{n}}_{ph}$ $=$ 1.0 $\times$ 
10$^{48}$ s$^{-1}$ as before while ${\dot{n}}_{ph}$ $=$ 5.0 $\times$ 10$^{49}$ s$^{-1}$ in the other set.  
This larger rate sets R$_{\omega}$ $=$ 0.26 pc, outside the central plateau.  One calculation in each set 
had a cutoff temperature of 1000 K while the other had a cutoff of 100 K; the background temperature in 
each was 100 K.  The evolution of the front was followed for 1.3 Myr in the first set and for 45 kyr in 
the second set (much less time is required when there is rapid escape down the gradient).  Flow profiles 
for all four calculations appear in Fig \ref{fig:r2}.
 
Even out to 20 pc, ten times the distance to which our earlier models were followed, the density and 
ionization fraction profiles in the left panel of Fig \ref{fig:r2} reveal that the I-front remains trapped 
by the shock at the lower photon rates.  The 100 K shell is only half as thick as the 1000 K shell but has 
ten times its density.  The higher recombination rate in the cooler shell offsets its smaller width to 
confine the front.  Note that radiative cooling in both cases lowers the shell to the cutoff temperature.  
When we extend the grid boundary to 200 pc at the same spatial resolution (thereby ensuring that the front 
remains on the grid for the 10 Myr lifetime of a star with these emission rates \citep{s02}), we find that 
the front finally breaks through the 1000 K shell at 90 pc and through the 100 K shell at 64 pc.  In these
calculations the shock gains strength as it descends the density gradient.  Line cooling still collapses gas
into a relatively thin layer at the base of the shell but cannot cool all the material that accumulates as 
the shock accelerates so the shell heats and thickens.  When the shell's outer layers reach 10,000 K, 
collisional ionizations assist the I-front to break through the shock.  

On the other hand, the right panel demonstrates that if R$_S$ $=$ 0.26 pc $>$ r$_c$, the front breaks free  
of the shock just past R$_S$ as predicted by \citet{ftb90}.  Reverting to R-type, the front quickly exits 
the spherical cloud, leaving behind a steepening ionized core shock at the site of detachment as seen in 
the density panel.  Little gas dynamical evolution occurs over the time in which the front crosses the grid.  
The higher density of the cooler shell slightly delays I-front breakout (by $\sim$ 2 - 3 kyr).  This implied 
that instability growth would be particularly explosive in cases where R$_S$ lies beyond the core radius.  

\subsection{S31 Breakout Instability}

To test this conjecture we devised a new S31 run with ${\dot{n}}_{ph}$ $=$ 5.0 $\times$ 10$^{49}$ s$^{-1}$.  
We imposed random variations on the density as before but beyond radii of 0.29 pc in order to allow the 
I-front to become fully D-type at R$_{\omega}$ = 0.26 pc before encountering the perturbations.  Temperature
and density images for $\phi$ = $\pi$/2 in this calculation are shown in Fig (\ref{fig:S31brk}) for 9.4, 13.2, 
and 15.1 kyr.  Radiation breakout through cracks in the shock rapidly ensues and the front erupts through the 
envelope in spikes which are R-type, as evidenced by the short escape times and the lack of hydrodynamic 
response in the ionized fingers.  The slightly jagged edges in some of the spikes are artifacts of the 
spherical polar coordinate mapping to cartesian coordinates.  Fragmentation of the shell is visible at 
early times as bright dots in the shell in the density panels but it soon disappears as the clumps are 
photoionized and then dispersed on time scales setby the sound speed in the postfront gas.  The maximum  
density achieved in this simulation was 5.47 $\times$ 10$^4$ cm$^{-3}$.  The spikes remain fairly narrow 
because they elongate rapidly in comparison to hydrodynamical time scales, but they do begin to drive weak 
transverse shocks at their base as the ionized gas begins to laterally expand.  These unstable modes 
do not fall in any of the categories discussed so far because they evolve as fast R-type structures.

\section{R-Type Shadow Instabilities}

In this test, a plane-parallel R-type ionization front enters the yz-face of a rectangular box with a 
uniform gas density except for a spherical bump slightly offset from the face of entry.  The density in 
the bump varies linearly in radius from the ambient value at its center to either 50\% below or above this 
value at its surface.  Our simulation volume has a length of 1.0 pc along the x-axis and 0.25 pc along the 
y and z-axes, with 1000, 250, and 250 zones in the x, y, and z directions respectively.  The gas number 
density in the box is 1000 cm$^{-3}$ and is hydrogen only.  The temperature of the gas was set to 72 K, 
yielding a sound speed c$_{s}$ of 1 km/s.  The fixed energy $\epsilon_{\Gamma}$ per photoionization was 
0.8 eV to yield initial postfront temperatures of 10$^4$.  The incident photon flux along the x-axis at 
the entry face was 3.0 $\times$ 10$^{11}$ cm$^{-2}$ s$^{-1}$ to approximate that of an O star at its 
Str\"{o}mgren radius.  The radius of the perturbation was 0.0125 pc and was positioned 0.05 pc along the 
x-axis and centered in the yz-plane.  Reflecting and outflow boundary conditions were applied at the 0 pc 
and 1.0 pc faces, respectively, with reflecting boundaries along the other four faces.

Our physical parameters for this test are similar to those of the two-dimensional simulations of \citet{rjw99}, 
except for cooling.  In those models near-isothermality was enforced by setting $\gamma$ = 1.1 and the internal 
energy of the ionized gas to be constant to guarantee c$_s$ $\sim$ 10 km/sec.  In contrast, we applied the same 
cooling algorithm described earlier with a cutoff temperature of 1000 K.  The front reaches a Str\"{o}mgren 
distance of 0.32 pc at approximately 1000 yr: 
\begin{equation}
x_{Str} = \displaystyle{\frac{F_{0}}{{n}^{2}{\alpha}_{B}}}
\end{equation}
where F$_{0}$ is the photon flux (cm$^{-2}$ s$^{-1}$), n is the number density of neutral hydrogen and 
${\alpha}_B$ is the case B recombination coefficient.  The simulation volume was partitioned into 25
uniform tiles, with five divisions along both the y-axis and z-axis and each tile spanning the length
of the volume in the x direction.  

\begin{figure*}
\epsscale{1.0}
\plotone{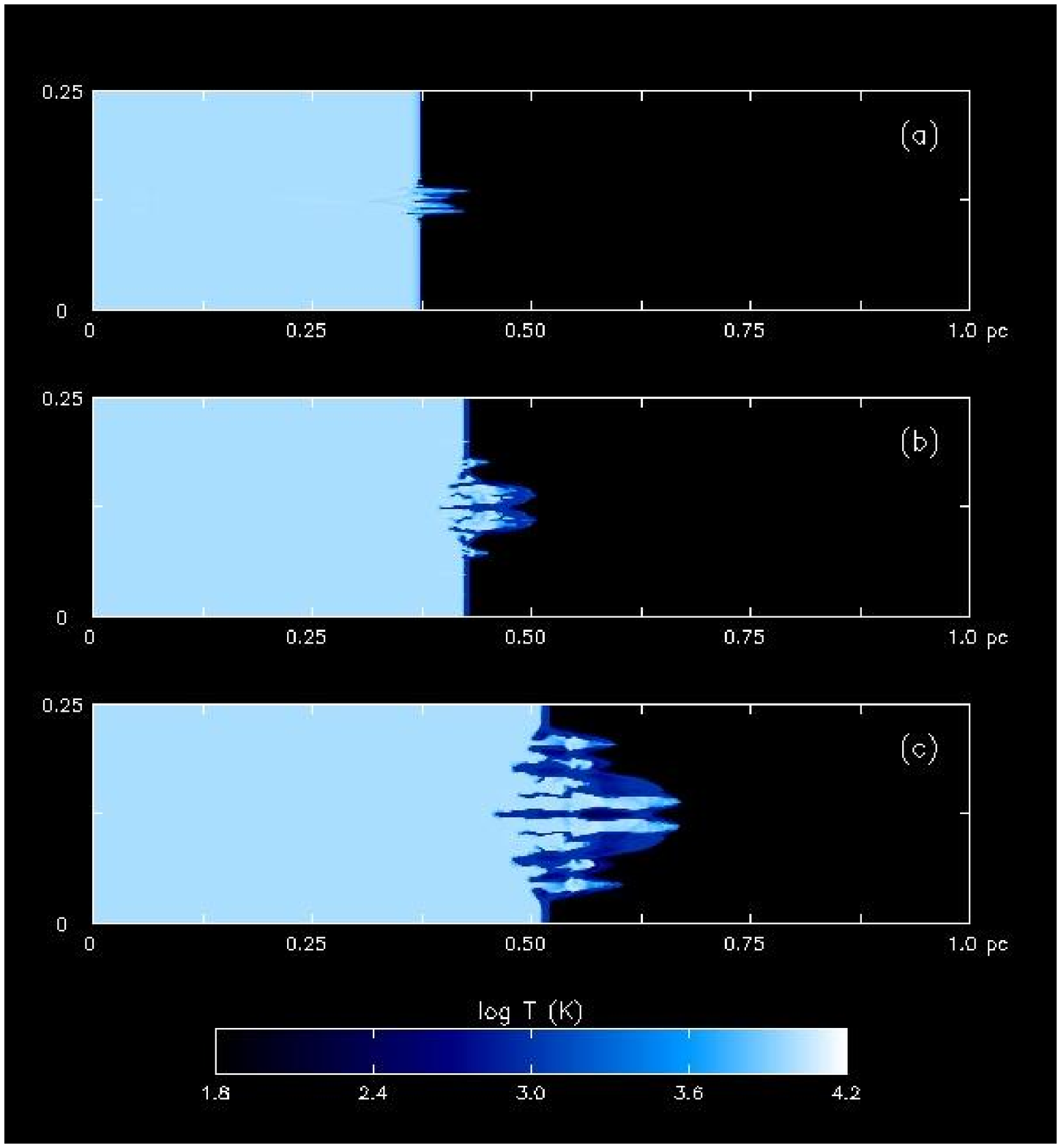}\vspace{0.15in}
\caption{Temperature evolution of an I-front shadow instability due to an underdense spherical 
underdensity: (a) 1.9 kyr, (b) 4.9 kyr, and (c) 12.6 kyr.\label{fig: rwmDM_temp}} 
\vspace{0.075in}
\end{figure*}

\begin{figure*}
\epsscale{1.0}
\plotone{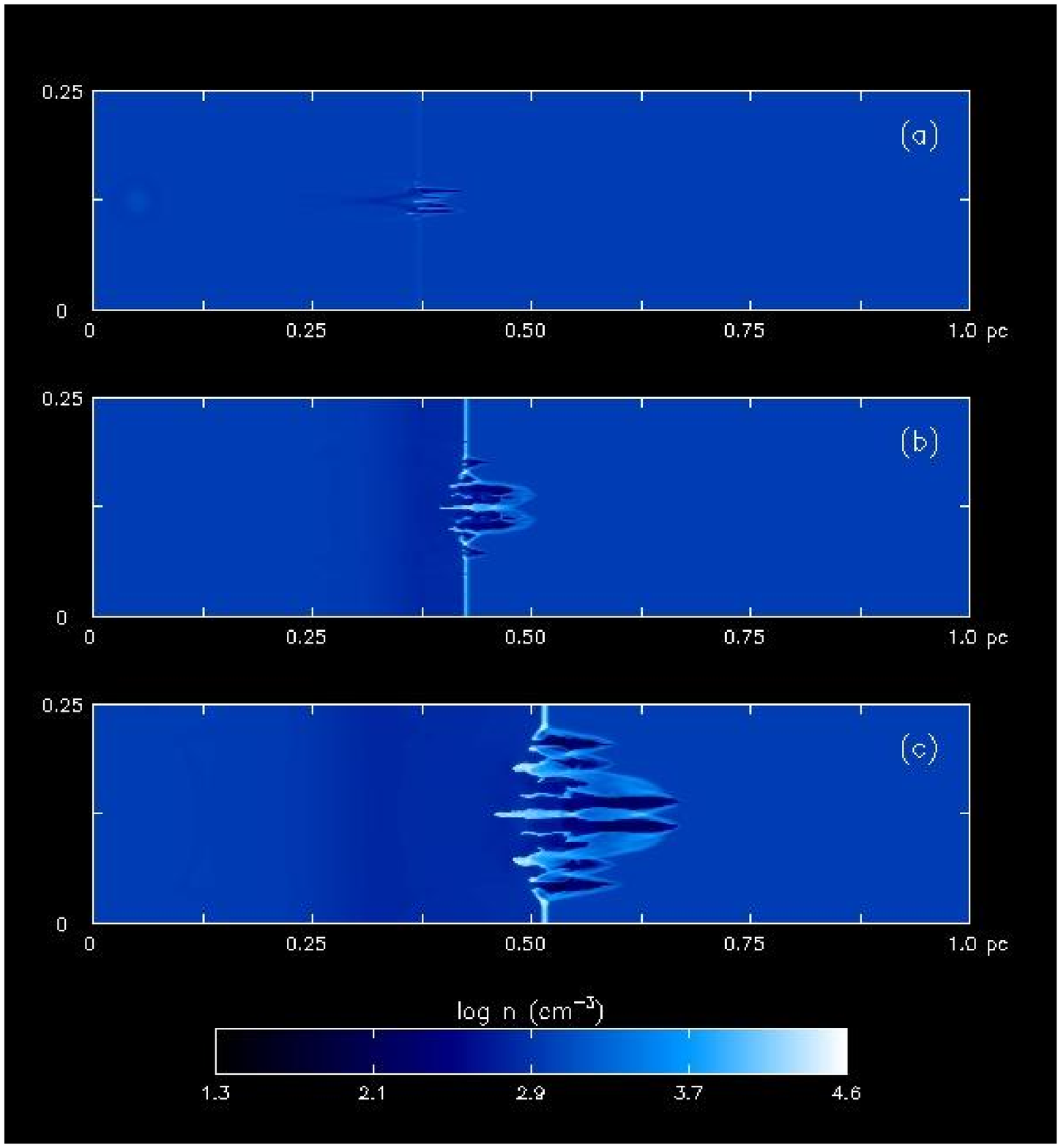}\vspace{0.15in}
\caption{Density evolution of an I-front shadow instability due to the spherical underdensity: (a) 
1.9 kyr, (b) 4.9 kyr, and (c) 12.6 kyr. \label{fig: rwmDM_dens}} 
\vspace{0.075in}
\end{figure*}

We show xy slices at z = 0 of the temperature and density evolution of the shadowing instability in 
Figs \ref{fig: rwmDM_temp} and \ref{fig: rwmDM_dens} due to an underdense perturbation at 1.9 kyr, 
4.9 kyr, and 12.6 kyr.  At 1.9 kyr the front has transitioned to D-type and the original corrugation 
in the R-type front is well into the nonlinear growth phase.  The expanding photoevaporated perturbation 
is faintly visible on the left of the figure.  The twin peaks in the instability are xy slices through 
a ringed jet that is due to the radial density profile of the spherical bump:  the R-type front 
preferentially advances along the lines of sight parallel to the x-axis that cut the underdense regions 
close to the surface of the bump.  Along the line of sight piercing the bump through its center, the front 
advances at nearly the same rate as in the unperturbed medium because the densities along this path through 
the bump are close to those in the surrounding gas.  

At later times cooling in the shell fractures the instability into multiple jets with highly nonlinear 
growth and morphologies reminiscent of the density structures visible in the two-dimensional numerical 
experiments.  No temperature images of the 1999 work are available so it is not possible to directly
compare the distribution of ionized and shocked gas within the jets or verify the existence of the 
ionized extensions into the shock that are present in our models.  The unperturbed regions of the front 
snowplow gas into a somewhat broader shell than in the \citet{rjw99} models because the gas can only cool 
to 1000 K.  The length of the jet in our model (0.14 pc) is somewhat shorter than in the Williams runs 
due in part to the lower ionizing flux.  However, intrinsic differences between two and three-dimensional 
flows may also come into play, as the additional degree of freedom opens transverse flow channels that may
quench the instability.  Indeed, some vorticity is evident in the flow at later times, a possible signature 
that instability growth is coupling to turbulent motion in the gas and being reduced.  The densities in the 
jets range from 19.0 cm$^{-3}$ to 3.72 $\times$ 10$^4$ cm$^{-3}$, a factor of 4 or 5 lower than in the 1999 
experiments.  The greater maxima in the earlier work in part arise from the higher density jumps permitted 
across the isothermal shocks in those models as well as from the lower temperatures to which the postshock 
gas can cool.  As the central jet advances it widens as ionized gas in its interior expands perpendiculary
to the flow.

Two features also worth mention are the density knots in the shock located above and below the midplane 
of the jets at their base at 1.9 kyr.  As the front advances, radiation escapes past the knots and forms
jets that migrate across the face of the shock as they elongate.  Shortly thereafter, smaller disturbances
in the shock materialize further from the base of the central jet, possibly triggered by acoustic waves
along the shock or perhaps merely numerical artifacts.  These density fluctuations quickly cool and create
cracks in the shock through which ionizing photons penetrate, forming smaller subsidiary jets that later
merge with the widening central jet in complicated hydrodynamical flows.  The original knots are present 
in the two-dimensional runs and are attributed by Williams to a type of zero-wavelength odd-even numerical 
instability (zero-wavelength in the sense that it would be manifest at all resolutions).  The phenomena 
apparently arise because the I-front itself is not resolved by the grid.  Just as postshock oscillations 
can come to dominate shocked flow not broadened by a few zones with artificial viscosity in lower-order 
hydrodynamical schemes, under certain circumstances sharp I-fronts can preferentially advance in an 
oscillatory fashion across adjacent zones.  Williams removed this effect in his simulations by setting 
the photoionization cross section to be a function of zone width (in direct analogy to artificial viscosity), 
hence broadening the front.  Practical applications employing multifrequency radiative transfer (for 
Population III stellar and miniquasar hard spectra) may obviate this feature because the spread in photon
mean free paths would naturally widen I-fronts in many environments without recourse to artificial means.
   
\section{Conclusion}

For the past ten years, the general paradigm of ionization front instabilities in the ISM has been one 
of swordlike structures with unrestrained amplitude growth in a wide variety of environments.  Our new 
models temper this view with saturated modes that likely drive turbulent flows in some instances.  The 
morphology of the S31 instability in particular suggests that some modes may efficiently drive turbulence 
in the shocked shell, with possible implications for time scales of subsequent star formation in the 
region.  The numerous short-wavelength ionized spikes in the shock at early times later gather into fewer 
long-wavelength structures, so energy is transported at first from smaller scales to larger ones by the 
flow.  Later, these large scale flows crumple into ever smaller scales as the saturated instabilities 
devolve into shear motions and develop the vorticity evident in the Kelvin-Helmholtz features appearing 
in panel (f) of Fig \ref{fig:S31DM}.  We expect this process to be more prominent in actual molecular 
clouds because the r$^{-2}$ profile of a core truncates in relatively high intra-cloud densities that 
would further flatten the instabilities.  While our current mesh can resolve the shell breakup and 
ionized fingers well, these features would have to be evolved for longer times on finer grids able to 
capture turbulent cascades to determine the extent to which turbulence arises in the cloud.  Nevertheless, 
the possibility that ionization front instabilities together with winds from massive stars and supernova 
explosions may provide an important source of turbulent support in star forming clouds in the galaxy today 
is an intriguing one.   

We find that the dimensionality of our models distinguishes them from previous work more than their improved 
physics.  The structures emerging in our simulations bear resemblance to many objects visible in the ISM 
today.  The globules and pillars in the S31 run are reminiscent of those seen in the Eagle nebula while 
the long wispy extensions in the S22 run are comparable to those in many planetary nebula such as NGC 6751.  
The clumping that is prevalent in the galaxy is also evident in our models.  In reality, ionization front 
instabilities compete with wind-blown structures and blast waves to produce the panoply of features in the 
local interstellar medium.  How all three act in concert to create the circumstellar environments of massive 
stars and gamma-ray bursts is an important question to be addressed in future work.  Likewise, how H II 
region instabilities interact with the magnetic fields threading molecular clouds remains poorly understood.
We expect that unstable modes will not be seriously suppresed on parsec scales because of the relative
magnitudes of ionized gas and magnetic field pressures, but their evolution could be altered on larger
scales as they approach pressure equilibrium with their surroundings.

A question of relevance to cosmological reionization is whether I-front instabilities arise in the
primordial clouds of primeval massive stars and protogalaxies.  In pristine gas shocks must resort to 
atomic hydrogen line cooling and Compton cooling at high redshifts, which are far less efficient processes
than metal line emission.  If the shocked flows are non-radiating, no Vishniac instabilities can develop
or be amplified by radiation so the fronts would be stable.  On the other hand, I-front shocks accelerate 
if they emerge from cosmological minihalos with density gradients steeper than r$^{-2}$.  Their exit may 
thus be subject to Rayleigh-Taylor instabilities that ionizing radiation could blossom into much larger 
structures.  We further note that hard Population III or miniquasar UV and x-ray spectra can significantly 
broaden comological I-fronts, and it is unclear what impact the layer of partial ionization will have on 
instability formation.  These questions together with how I-fronts sculpt gas clouds in the early intergalactic 
medium will be the focus of studies to come.

Finally, we caution that all our models adopted the on-the-spot approximation, in which recombination 
photons are assumed to be reabsorbed in their zone of origin before escaping.  In reality, reprocessed
radiation isotropically emitted within the ionized fingers may halt their growth by eroding them from 
within.  Recombination radiation is believed to dominate the photon budget in the outer radii of centrally 
concentrated H II regions like those examined in our study \citep{r05}.  Direct transport of diffuse ionizing 
radiation exhausts current ray tracing techniques, so the implementation of flux-limited diffusion for 
recombination photons in ZEUS-MP is currently underway in order to study their influence on instabilities 
in future models.  Nevertheless, we expect that diffuse radiation will at most blunt the instabilities, not 
prevent their formation or destroy them because the fragmentation of the shell and breakout of the radiation 
through the cracks is a very robust feature of the fronts.

\acknowledgments

We thank the anonymous referee, whose insightful critique signicantly improved the quality of this paper.
DW is grateful for useful discussions concerning this work with Guillermo Garcia-Segura, Alex Heger, 
Mordecai-Mark Mac Low and Marcelo Alvarez, as well as for conversations with with Hajime Susa and Naoki 
Yoshida at the University of Washington INT workshop ``The First Stars and Evolution of the Early Universe'' 
held from June 19 to July 21, 2006.  He also thanks Brian O'Shea for his software package for mapping the 
spherical polar coordinate data in this study onto cartesian grids for the images in this paper.  This 
work was carried out under the auspices of the National Nuclear Security Administration of the U.S. 
Department of Energy at Los Alamos National Laboratory under Contract No. DE-AC52-06NA25396.  The 
simulations were performed at SDSC and NCSA under NRAC allocation MCA98N020.


\begin{thebibliography}{}




\bibitem[Arquilla \& Goldsmith(1985)]{ag85} Arquilla, R., \& Goldsmith, P.~F.\ 1985, 
\apj, 297, 436 
\bibitem[Axford(1964)]{ax64} Axford, W.~I.\ 1964, \apj, 140, 112 
\bibitem[Bertoldi(1989)]{brt89} Bertoldi, F.\ 1989, \apj, 346, 735 
\bibitem[Brand(1981)]{bnd81} Brand, P.~W.~J.~L.\ 1981, \mnras, 197, 217 
\bibitem[Canto et al.(1998)]{cet98} Canto, J., Raga, A., Steffen, W., \& Shapiro, P.\ 1998, 
\apj, 502, 695 
\bibitem[Capriotti(1973)]{cap73} Capriotti, E.~R.\ 1973, \apj, 179, 495 
\bibitem[Dale et al.(2007)]{det07} Dale, J.~E., Bonnell, I.~A., \& Whitworth, A.~P.\ 2007, 
\mnras, 42 
\bibitem[Dalgarno \& McCray(1972)]{dm72} Dalgarno, A. \& McCray, R.~A. \ 1972, \araa, 10, 375
\bibitem[Franco et al.(1998)]{fet98} Franco, J., Diaz-Miller, R.~L., Freyer, T., \& 
Garcia-Segura, C.\ 1998, ASP Conf.~Ser.~141: Astrophysics From Antarctica, 141, 154 
\bibitem[Franco \etal(1990)]{ftb90} Franco, J., Tenorio-Tagle, G., \& Bodenheimer, P.\ 1990, 
\apj, 349, 126 
\bibitem[Freyer et al.(2003)]{fhy03} Freyer, T., Hensler, G., \& Yorke, H.~W.\ 2003, \apj, 594, 
888 
\bibitem[Frieman(1954)]{f54} Frieman, E.~A.\ 1954, \apj, 120, 18 
\bibitem[Fuller \& Couchman(2000)]{fc00} Fuller, T.~M.~\& Couchman, H.~M.~P.\ 2000, 
\apj, 544, 6 
\bibitem[Garcia-Segura \& Franco(1996)]{gsf96} Garcia-Segura, G., \& Franco, J.\ 1996, 
\apj, 469, 171 
\bibitem[Giuliani(1979)]{gu79} Giuliani, J.~L.\ 1979, \apj, 233, 280 
\bibitem[Gregorio Hetem et al.(1988)]{het88} Gregorio Hetem, J.~C., Sanzovo, G.~C., \& Lepine, 
J.~R.~D.\ 1988, \aaps, 76, 347 
\bibitem[Hayes \etal(2006)]{jch06} Hayes, J.~C., Norman, M.~L., Fiedler, R.~A., Bordener, J.~O., 
Li, P.~S., Clark, S.~E., ud-Doula, A., Mac Low, M.~-M., astro-ph/0511545. 
\bibitem[Kahn(1958)]{kh58} Kahn, F.~D.\ 1958, Reviews of Modern Physics, 30, 1058 
\bibitem[Mac Low \& Norman(1993)]{mn93} Mac Low, M.-M., \& Norman, M.~L.\ 1993, \apj, 407, 207 
\bibitem[Mellema et al.(1998)]{met98} Mellema, G., Raga, A.~C., Canto, J., Lundqvist, P., Balick, 
B., Steffen, W., \& Noriega-Crespo, A.\ 1998, \aap, 331, 335 
\bibitem[Mizuta et al.(2005)]{miz05} Mizuta, A., Kane, J.~O., Pound, M.~W., Remington, B.~A., 
Ryutov, D.~D., \& Takabe, H.\ 2005, \apj, 621, 803 
\bibitem[Mizuta et al.(2006)]{miz06} Mizuta, A., Kane, J.~O., Pound, M.~W., Remington, B.~A., 
Ryutov, D.~D., \& Takabe, H.\ 2006, \apj, 647, 1151 
\bibitem[Newman \& Axford(1967)]{na67} Newman, R.~C., \& Axford, W.~I.\ 1967, \apj, 149, 571 
\bibitem[Osterbrock(1989)]{o89} Osterbrock, D. Astrophysics of Gaseous Nebulae and
Active Galactic Nuclei, University Science Books 1989.
\bibitem[Ritzerveld(2005)]{r05} Ritzerveld, J.\ 2005, \aap, 439, L23 
\bibitem[Schaerer(2002)]{s02} Schaerer, D.\ 2002, \aap, 382, 28
\bibitem[Shapiro et al.(2004)]{shp04} Shapiro, P.~R., Iliev, I.~T., \& Raga, A.~C.\ 2004, \mnras, 
348, 753 
\bibitem[Soker(1998)]{sok98} Soker, N.\ 1998, \mnras, 299, 562 
\bibitem[Spitzer(1954)]{sp54} Spitzer, L.~J.\ 1954, \apj, 120, 1 
\bibitem[Sysoev(1997)]{sy97} Sysoev, N.~E.\ 1997, Astronomy Letters, 23, 409 
\bibitem[Vandervoort(1962)]{v62} Vandervoort, P.~O.\ 1962, \apj, 135, 212 
\bibitem[Vishniac(1983)]{v83} Vishniac, E.~T.\ 1983, \apj, 274, 152 
\bibitem[Whalen \etal(2004)]{wan04} Whalen, D., Abel, T., \& Norman, M.~L.\ 2004, \apj, 610, 14 
\bibitem[Whalen \& Norman(2006)]{wn06} Whalen, D., \& Norman, M.~L.\ 2006, \apjs, 162, 281 
\bibitem[Williams(1999)]{rjw99} Williams, R.~J.~R.\ 1999, \mnras, 310, 789 
\bibitem[Williams(2002)]{rjw02} Williams, R.~J.~R.\ 2002, \mnras, 331, 693 
\bibitem[Williams et al.(2000)]{rjw00} Williams, R.~J.~R., Dyson, J.~E., \& Hartquist, T.~W.\ 
2000, \mnras, 314, 315 
\bibitem[Yoshida et al.(2006)]{yet06} Yoshida, N., Oh, S.~P., Kitayama, T., \& Hernquist, L.\ 
2006, astro-ph/0610819 


\end{thebibliography}
\end{document}